\documentclass[10pt,a4paper,onecolumn]{IEEEtran}
\usepackage[lmargin=2cm, rmargin=2cm, tmargin=2cm, bmargin=2cm]{geometry}
\usepackage{ccaption}
\usepackage[T1]{fontenc}
\usepackage{graphicx}
\usepackage{fancyhdr}
\usepackage{helvet}
\usepackage{floatflt}
\usepackage{colortbl}
\usepackage{fancyvrb}
\usepackage{wrapfig}
\usepackage[round,authoryear]{natbib}
\usepackage[latin1]{inputenc}
\usepackage[T1]{fontenc}
\usepackage[english]{babel}
\usepackage{graphicx}
\usepackage{subfigure}
\usepackage{amsfonts}
\usepackage{amsmath}
\usepackage{theorem}
\usepackage{ae}

\newtheorem{assumption}{Assumption}

\newtheorem{theorem}{Theorem}

\title{Adaptive fuzzy control of electrohydraulic servosystems}
\author{Wallace Moreira Bessa, Max Suell Dutra, Edwin Kreuzer}
\date{}

\begin{document}

\maketitle

\abstract{Electrohydraulic servosystems are widely employed in industrial applications such as robotic manipulators, 
active suspensions, precision machine tools and aerospace systems. They provide many advantages over 
electric motors, including high force to weight ratio, fast response time and compact size. However, 
precise control of electrohydraulic actuated systems, due to their inherent nonlinear characteristics, 
cannot be easily obtained with conventional linear controllers. Most flow control valves can also 
exhibit some hard nonlinearities such as dead-zone due to valve spool overlap. This work describes 
the development of an adaptive fuzzy controller for electrohydraulic actuated systems with unknown 
dead-zone. The stability properties of the closed-loop systems was proven using Lyapunov stability 
theory and Barbalat's lemma. Numerical results are presented in order to demonstrate the control 
system performance.}

\section{Introduction}

Electrohydraulic actuators play an essential role in several branches of industrial activity and are 
frequently the most suitable choice for systems that require large forces at high speeds. Their 
application scope ranges from robotic manipulators to aerospace systems. Another great advantage of 
hydraulic systems is the ability to keep up the load capacity, which in the case of electric actuators 
is limited due to excessive heat generation. 

However, the dynamic behavior of electrohydraulic systems is highly nonlinear, which in fact makes 
the design of controllers for such systems a challenge for the conventional and well established linear 
control methodologies. The increasing number of works dealing with control approaches based on modern
techniques shows the great interest of the engineering community, both in academia and industry, in 
this particular field. The most common approaches are the adaptive \citep{knohl1,yao1} and variable
structure \citep{bonchis1,jerouane1,liu1,mihajlov1} methodologies, but nonlinear controllers based
on quantitative feedback theory \citep{sohl1,niksefat1}, optimal tuning PID control \citep{liu2}, 
integrator backstepping method \citep{chen2} and fuzzy model reference learning control \citep{testi1} 
were also presented.

In addition to the common nonlinearities that originate from the compressibility of the hydraulic 
fluid and valve flow-pressure properties, most electrohydraulic systems are also subjected to hard 
nonlinearities such as dead-zone due to valve spool overlap. It is well-known that the presence of a
dead-zone can lead to performance degradation of the controller and limit cycles or even instability
in the closed-loop system. 

Intelligent control has proven to be a very attractive approach to cope with uncertain nonlinear systems 
\citep{cobem2005,Bessa2017,Bessa2018,Bessa2019,Deodato2019,Lima2018,Lima2020,Tanaka2013}. 
By combining nonlinear control techniques, such as feedback linearization or sliding modes, with adaptive intelligent algorithms, 
for example fuzzy logic or artificial neural networks, the resulting intelligent control strategies can deal with the nonlinear 
characteristics as well as with modeling imprecisions and external disturbances that can arise.

In this work, an adaptive fuzzy controller is developed for electrohydraulic actuated systems with 
unknown dead-zone to deal with the position trajectory tracking problem. The adopted approach is
based on a recently proposed strategy \citep{ermac2005}, that does not requires previous knowledge
of dead-zone parameters. The global stability of the closed-loop system was proven using Lyapunov 
stability theory and Barbalat's lemma. Some numerical results are also presented in order to 
demonstrate the control system performance.

\section{Electrohydraulic system model}

In order to design the adaptive fuzzy controller, a mathematical model that represents the hydraulic 
system dynamics is needed. Dynamic models for such systems are well documented in the literature
\citep{merritt1,walters1}.

\begin{figure}[htb]
\centering
\includegraphics[width=0.8\textwidth]{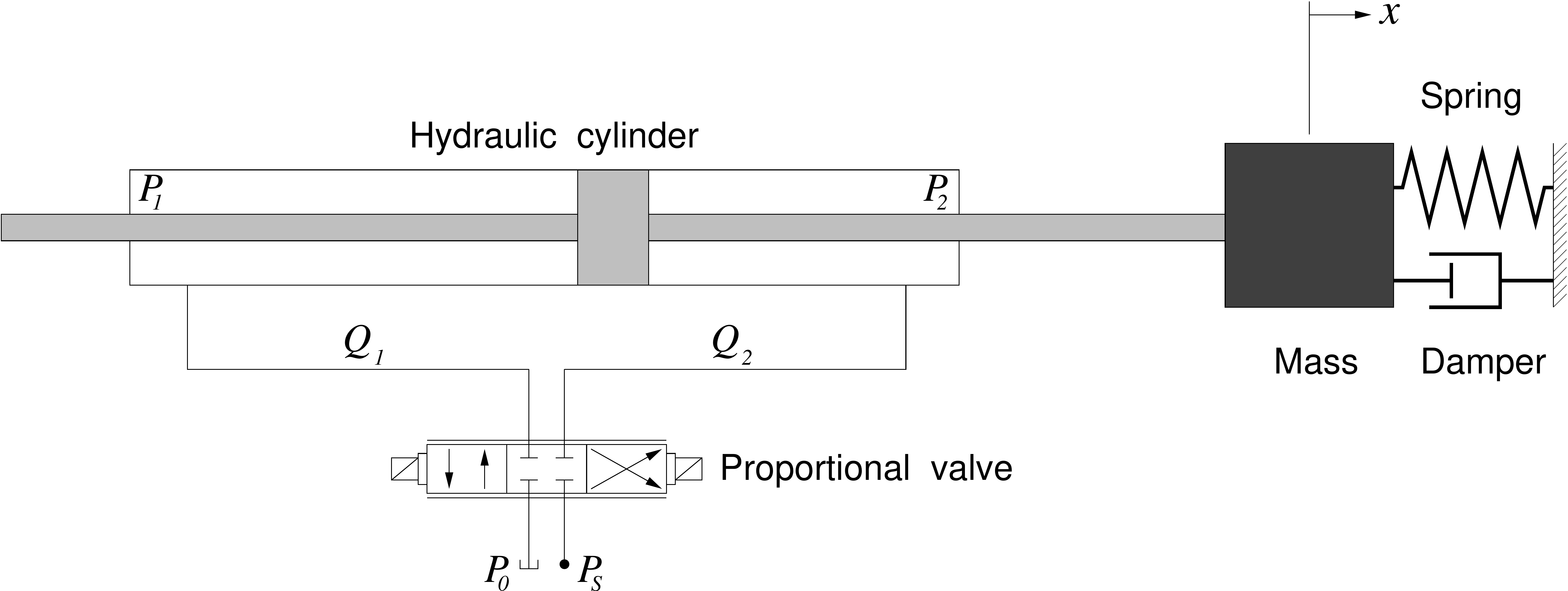}
\caption{Schematic diagram of the electrohydraulic servosystem.}
\label{fig:sistema}
\end{figure}

The electrohydraulic system considered in this work consists of a four-way proportional valve, 
a hydraulic cylinder and variable load force. The variable load force is represented by a 
mass--spring--damper system. The schematic diagram of the system under study is presented in
Fig.~(\ref{fig:sistema}).

The balance of forces on the piston leads to the following equation of motion: 

\begin{equation}
F_g=A_1P_1-A_2P_2=M_t\ddot{x}+B_p\dot{x}+K{x}
\label{eq:mov_aux}
\end{equation}

\noindent
where $F_g$ is the force generated by the piston, $P_1$ and $P_2$ are the pressures at each side of 
cylinder chamber, $A_1$ and $A_2$ are the ram areas of the two chambers, $M_t$ is the total mass of
piston and load referred to piston, $B_p$ is the viscous damping coefficient of piston and load, $K$
is the load spring constant and $x$ is the piston displacement.

Defining the pressure drop across the load as $P_L=P_1-P_2$ and considering that for a symmetrical
cylinder $A_p=A_1=A_2$, Eq.~(\ref{eq:mov_aux}) can be rewritten as

\begin{equation}
M_t\ddot{x}+B_p\dot{x}+K{x}=A_pP_L
\label{eq:mov}
\end{equation}

Applying continuity equation to the fluid flow, the following equation is obtained:

\begin{equation}
Q_L=A_p\dot{x}+C_{tp}+\frac{V_t}{4\beta_e}\dot{P}_L
\label{eq:cont}
\end{equation}

\noindent
where $Q_L=(q_1+q_2)/2$ is the load flow, $C_{tp}$ the total leakage coefficient of piston, $V_t$ the 
total volume under compression in both chambers and $\beta_e$ the effective bulk modulus.

Considering that the return line pressure is usually much smaller than the other pressures involved
($P_0\approx0$) and assuming a closed center spool valve with matched and symmetrical orifices, the 
relationship between load pressure $P_L$ and load flow $Q_L$ can be described as follows

\begin{equation}
Q_L=C_dw\bar{x}_{sp}\sqrt{\frac{1}{\rho}\big(P_s-\mbox{sgn}(\bar{x}_{sp})P_L\big)}
\label{eq:fluxo}
\end{equation}

\noindent
where $C_d$ is the discharge coefficient, $w$ the valve orifice area gradient, $\bar{x}_{sp}$ the 
effective spool displacement from neutral, $\rho$ the hydraulic fluid density, $P_s$ the supply 
pressure and sgn($\cdot$) is defined by

\begin{equation}
\mbox{sgn}(z) = \left\{\begin{array}{rc}
-1&\mbox{if}\quad z<0 \\
0&\mbox{if}\quad z=0 \\
1&\mbox{if}\quad z>0
\end{array}\right.
\label{eq:fun_sgn}
\end{equation}

\begin{figure}[hbt]
\centering
\includegraphics[width=0.3\textwidth]{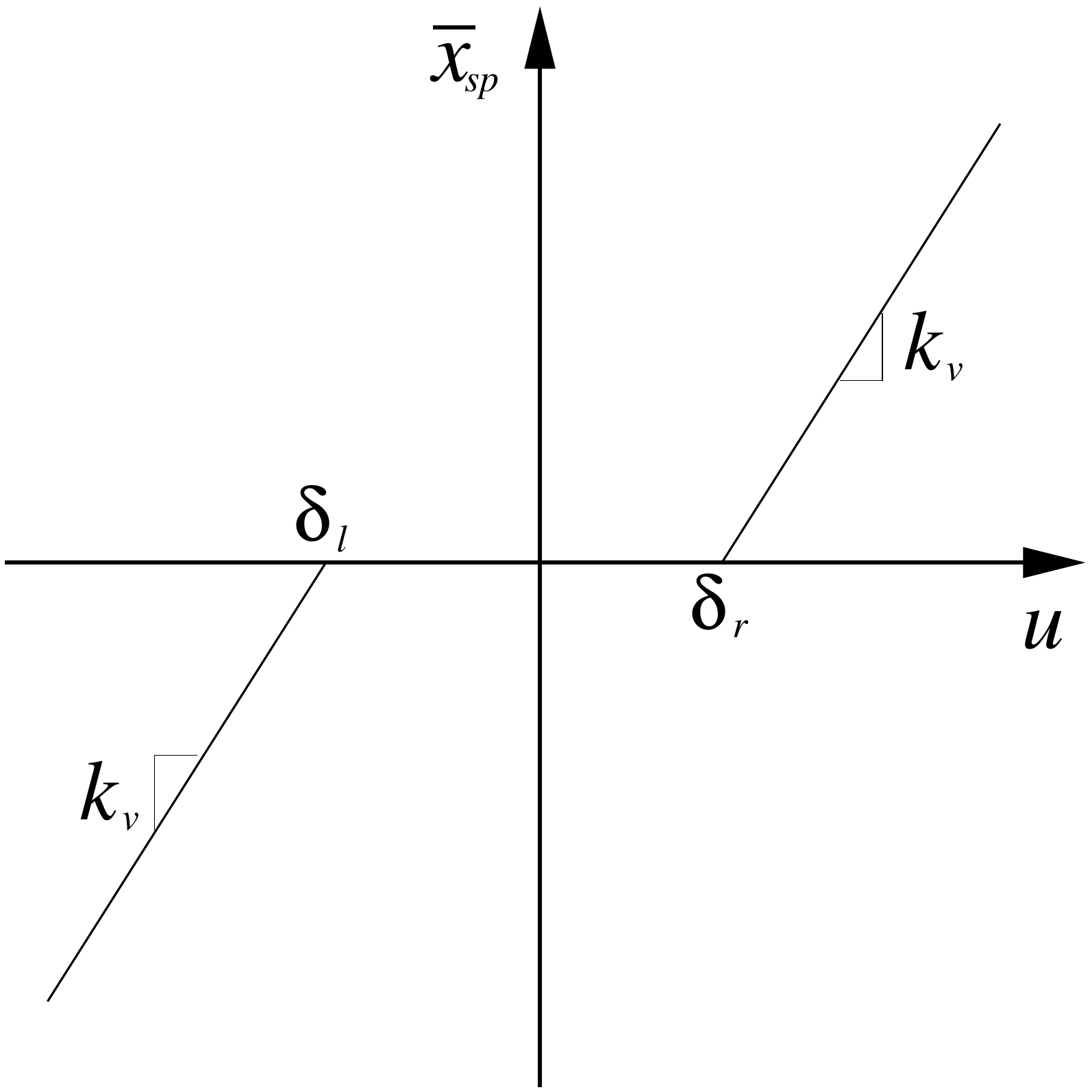} 
\caption{Dead-zone nonlinearity.}
\label{fig:dzone}
\end{figure}

Assuming that the dynamics of the valve are fast enough to be neglected, the valve spool displacement 
can be considered as proportional to the control voltage ($u$). For closed center valves, or even in the 
case of the so-called critical valves, the spool presents some overlap. This overlap prevents from leakage 
losses but leads to a dead-zone nonlinearity within the control voltage, as shown in Fig.~(\ref{fig:dzone}).

The dead-zone nonlinearity presented in Fig.~(\ref{fig:dzone}) can be mathematically described by:

\begin{equation}
\bar{x}_{sp}(t) = \left\{\begin{array}{ll}
k_v\,\big(u(t)-\delta_l\big)&\mbox{if}\quad u(t)\le\delta_l \\
0&\mbox{if}\quad  \delta_l<u(t)<\delta_r\\
k_v\,\big(u(t)-\delta_r\big)&\mbox{if}\quad u(t)\ge\delta_r
\end{array}\right.
\label{eq:dzone1}
\end{equation}

\noindent
where $k_v$ is the valve gain and the parameters $\delta_l$ and $\delta_r$ depends on the size of 
the overlap region.

For control purposes, as shown by \citet{ermac2005}, Eq.~(\ref{eq:dzone1}) can be rewritten in a more 
appropriate form:

\begin{equation}
\bar{x}_{sp}(t) = k_v\big(u(t)-d(u)\big)
\label{eq:dzone2}
\end{equation}

\noindent
where $d(u)$ can be obtained from Eq.~(\ref{eq:dzone1}) and Eq.~(\ref{eq:dzone2}): 

\begin{equation}
d(u) = \left\{\begin{array}{ll}
\delta_l&\mbox{if}\quad u(t)\le\delta_l \\
u(t)&\mbox{if}\quad  \delta_l<u(t)<\delta_r\\
\delta_r&\mbox{if}\quad u(t)\ge\delta_r
\end{array}\right.
\label{eq:dzone3}
\end{equation}

Combining equations (\ref{eq:mov}), (\ref{eq:cont}), (\ref{eq:fluxo}), (\ref{eq:dzone2}) and 
(\ref{eq:dzone3}) leads to a third-order differential equation that represents the dynamic behavior
of the electrohydraulic system:

\begin{equation}
\dddot{x}=-\mathbf{a^\mathrm{T}x}+bu-bd(u)
\label{eq:modelo}
\end{equation}

\noindent
where $\mathbf{x}=[x,\dot{x},\ddot{x}]^\mathrm{T}$ is the state vector with an associated coefficient
vector $\mathbf{a}=[a_0,a_1,a_2]^\mathrm{T}$ defined according to

\begin{displaymath}
\displaystyle
a_0=\frac{4\beta_eC_{tp}K}{V_tM_t}\:\:\:\:\:\:;\:\:\:\:\:\:
a_1=\frac{K}{M_t}+\frac{4\beta_eA_{p}^2}{V_tM_t}+\frac{4\beta_eC_{tp}B_p}{V_tM_t}\:\:\:\:\:\:;\:\:\:\:\:\: 
a_2=\frac{B_p}{M_t}+\frac{4\beta_eC_{tp}}{V_t}
\end{displaymath}

\noindent
and

\begin{displaymath}
\displaystyle
b=\frac{4\beta_eA_p}{V_tM_t}C_dwk_v\sqrt{\frac{1}{\rho}\big[P_s-\mbox{sgn}(u)\big(M_t\ddot{x}+B_p\dot{x}
+K{x}\big)/A_p\big]}
\end{displaymath}

Based on the dynamic model presented in Eq.~(\ref{eq:modelo}), an adaptive fuzzy controller will be
developed in the next section.

\section{Adaptive fuzzy controller}

Consider the trajectory tracking problem and let $\mathbf{\tilde{x}}=\mathbf{x}-\mathbf{x}_d=[\tilde{x},
\dot{\tilde{x}},\ddot{\tilde{x}},\dddot{\tilde{x}}]^\mathrm{T}$ be the tracking error associated to a 
desired trajectory $\mathbf{x}_d=[x_d,\dot{x}_d,\ddot{x}_d,\dddot{x}_d]^\mathrm{T}$. 

Now, defining a combined tracking error measure $e=\mathbf{c^\mathrm{T}\tilde{x}}$, where $\mathbf{c}
=[c_0,c_1,1]^\mathrm{T}$ and the coefficients $c_0$ and $c_1$ chosen in order to make $p^2+c_1p+c_0$ 
a Hurwitz polynomial, the following control law can be proposed:

\begin{equation}
u=b^{-1}(\mathbf{a^\mathrm{T}x}+\dddot{x}_d-c_1\ddot{\tilde{x}}-c_0\dot{\tilde{x}})+\hat{d}(\hat{u})-\kappa e
\label{eq:lei}
\end{equation}

\noindent
where $\kappa$ is a strictly positive constant and $\hat{d}(\hat{u})$ an estimate of $d(u)$, that will 
be computed in terms of the equivalent control $\hat{u}=b^{-1}(\mathbf{a^\mathrm{T}x}+\dddot{x}_d-c_1
\ddot{\tilde{x}}-c_0\dot{\tilde{x}})$ by an adaptive fuzzy algorithm.

The adopted fuzzy inference system was the zero order TSK (Takagi--Sugeno--Kang), whose rules can be stated
in a linguistic manner as follows:

\begin{center}
\textit{If $\hat{u}$ is $\hat{U}_r$ then $\hat{d}_r=\hat{D}_r$}\hspace*{5pt};\hspace*{5pt}$r=1,2,\cdots,N$ 
\end{center}

\noindent
where $\hat{U}_r$ are fuzzy sets, whose membership functions could be properly chosen, and $\hat{D}_r$ is 
the output value of each one of the $N$ fuzzy rules.

Considering that each rule defines a numerical value as output $\hat{D}_r$, the final output $\hat{d}$ 
can be computed by a weighted average: 

\begin{equation}
\hat{d}(\hat{u}) = \frac{\sum_{r=1}^{N} w_r \cdot \: \hat{d}_r}{\sum_{r=1}^{N} w_r}
\label{eq:dc_media}
\end{equation}

\noindent
or, similarly,

\begin{equation}
\hat{d}(\hat{u}) = \mathbf{\hat{D}}^{\mathrm{T}}\mathbf{\Psi}(\hat{u})
\label{eq:dc_vetor}
\end{equation}

\noindent
where, $\mathbf{\hat{D}}=[\hat{D}_1,\hat{D}_2,\dots,\hat{D}_N]^{\mathrm{T}}$ is the vector containing the 
attributed values $\hat{D}_r$ to each rule $r$, $\mathbf{\Psi}(\hat{u})=[\psi_1(\hat{u}),\psi_2(\hat{u}), 
\dots,\psi_N(\hat{u})]^{\mathrm{T}}$ is a vector with components $\psi_r(\hat{u})= w_r/\sum_{r=1}^{N}w_r$ 
and $w_r$ is the firing strength of each rule.

To ensure the best possible estimate $\hat{d}(\hat{u})$, the vector of adjustable parameters can be 
automatically updated by the following adaptation law:

\begin{equation}
\mathbf{\dot{\hat{D}}}=-\varphi e \mathbf{\Psi}(\hat{u})
\label{eq:adapta}
\end{equation}

\noindent
where $\varphi$ is a strictly positive constant related to the adaptation rate. 

Before proving the closed-loop system stability, the following assumptions must be made:

\begin{assumption}
The states $x$, $\dot{x}$ and $\ddot{x}$ are available.
\label{th:dispon}
\end{assumption}
\begin{assumption}
The desired trajectory $x_d$ is $C^2$. Furthermore $x_d$, $\dot{x}_d$, $\ddot{x}_d$ and $\dddot{x}_d$ 
are available and with known bounds.
\label{th:xdbound}
\end{assumption}

\begin{theorem}
\label{th:teo1}
Let the electrohydraulic servosystem with a dead-zone (\ref{eq:dzone2})--(\ref{eq:dzone3}) at the input
be represented by Eq.~(\ref{eq:modelo}). Then, subject to Assumptions \ref{th:dispon}--\ref{th:xdbound}, 
the adaptive fuzzy controller defined by (\ref{eq:lei}), (\ref{eq:dc_vetor}) and (\ref{eq:adapta}) ensures 
the global stability of the closed-loop system and trajectory tracking.
\end{theorem}

\noindent
{\bf Proof:} Let a positive definite Lyapunov function candidate $V$ be defined as

\begin{equation}
V(t)=\frac{1}{2}e^2+\frac{b}{2\varphi}\mathbf{\Delta}^{\mathrm{T}}\mathbf{\Delta}
\label{eq:liap}
\end{equation}

\noindent
where $\mathbf{\Delta}=\mathbf{\hat{D}}-\mathbf{\hat{D}}^*$ and $\mathbf{\hat{D}}^*$ is the optimal
parameter vector, associated to the optimal estimate $\hat{d}^*(\hat{u})=d(u)$.

\noindent
Thus, the time derivative of $V$ is

\begin{displaymath}
\begin{array}{rcl}
\dot{V}(t) & = & e\dot{e}+b\varphi^{-1}\mathbf{\Delta^\mathrm{T}\dot{\Delta}}\\
& = & (\dddot{\tilde{x}}+c_1\ddot{\tilde{x}}+c_0\dot{\tilde{x}})e
+b\varphi^{-1}\mathbf{\Delta^\mathrm{T}\dot{\Delta}}\\
& = &\big(-\mathbf{a^\mathrm{T}x}+bu-bd(u)-\dddot{x}_d+c_1\ddot{\tilde{x}}+c_0\dot{\tilde{x}}\big)e
+b\varphi^{-1}\mathbf{\Delta^\mathrm{T}\dot{\Delta}}\\
\end{array}
\end{displaymath}

\noindent
Applying the proposed control law (\ref{eq:lei}) and noting that $\mathbf{\dot{\Delta}}=\mathbf{\dot{\hat{D}}}$, 
then

\begin{displaymath}
\begin{array}{rcl}
\dot{V}(t) & = & \big[b(\hat{d}-d)-\kappa e\big]e+b\varphi^{-1}\mathbf{\Delta^\mathrm{T}\dot{\hat{D}}}\\
& = & \big[b\mathbf{\Delta}^{\mathrm{T}}\mathbf{\Psi}(\hat{u})-\kappa e\big]e+b\varphi^{-1}\mathbf{
\Delta^\mathrm{T}\dot{\hat{D}}}\\
&=&-\kappa e^2+b\varphi^{-1}\mathbf{\Delta}^\mathrm{T}\big(\mathbf{\dot{\hat{D}}}-\varphi e\mathbf{\Psi}
(\hat{u})\big)\\
\end{array}
\end{displaymath}

\noindent
Furthermore, defining $\mathbf{\dot{\hat{D}}}$ according to (\ref{eq:adapta}), $\dot{V}(t)$ becomes

\begin{equation}
\dot{V}(t)=-\kappa e^2
\label{eq:liap_p}
\end{equation}

\noindent
which implies that $V(t)\le V(0)$ and that $e$ and $\mathbf{\Delta}$ are bounded. From the definition of 
$e$ and considering Assumption~\ref{th:xdbound}, it can be easily verified that $\dot{e}$ is also bounded.

\noindent
To establish the global stability of the closed loop system, the time derivative of $\dot{V}$ must be 
analyzed:

\begin{equation}
\ddot{V}(t)=-2\kappa e\dot{e}
\label{eq:liap_pp}
\end{equation}

\noindent
which implies that $\dot{V}(t)$ is also bounded and, from Barbalat's lemma, that $e\rightarrow0$ as
$t\rightarrow\infty$. 

\noindent
For $e=0$, the following error dynamics take place:

\begin{equation}
\ddot{\tilde{x}}+c_1\dot{\tilde{x}}+c_0\tilde{x}=0
\label{eq:hurwitz}
\end{equation}

\noindent
Thus, if the coefficients $c_0$ and $c_1$ were properly chosen, the associated characteristic
polynomial is a Hurwitz polynomial, which ensures the convergence of the tracking error to zero,
$\tilde{x}\rightarrow0$ as $t\rightarrow\infty$, and completes the proof.
\hfill$\square$ \vspace*{8pt}

In the following section some numerical simulations are presented in order to evaluate the performance 
of the adaptive fuzzy controller.

Some applications may require the use of robust controllers. In such a case, the reader is referred to 
the adaptive fuzzy sliding mode controllers presented in \citet{tese} and \citet{cnmac2005}. 
 
\section{Simulation results}

The simulation studies was performed with a numerical implementation, in C, with sampling rates of 400 Hz
for control system and 800 Hz for dynamic model. The adopted parameters for the electrohydraulic systems
were $P_s=7$ MPa, $\rho=850$ kg/m$^3$, $C_d = 0.6$, $w = 2.5\times10^{-2}$ m, $A_p=3\times10^{-4}$ m$^2$, 
$C_{tp}=2\times10^{-12}$ m$^3$/(s Pa), $\beta_e=700$ MPa, $V_t=6\times10^{-5}$ m$^3$, $M_t = 250$ kg,
$B_p = 100$ Ns/m, $K=75$ N/m, $\delta_l=-1.1$ V and $\delta_r=0.9$ V. The parameters of the controller
were $\lambda=8$, $\kappa=1$ and $\varphi=0.5$. For the fuzzy system were adopted triangular and trapezoidal 
membership functions for $\hat{U}_r$, with the central values defined as $C=\{-0.50\:;\:-0.10\:;\:-0.05\:;
\:0.00\:;\:0.05\:;\:0.10\:;\:0.50\}$. 

\begin{figure}[htb]
\centering
\mbox{
\subfigure[Tracking performance.]{\label{fig:graf1_1} 
\includegraphics[width=0.35\textwidth]{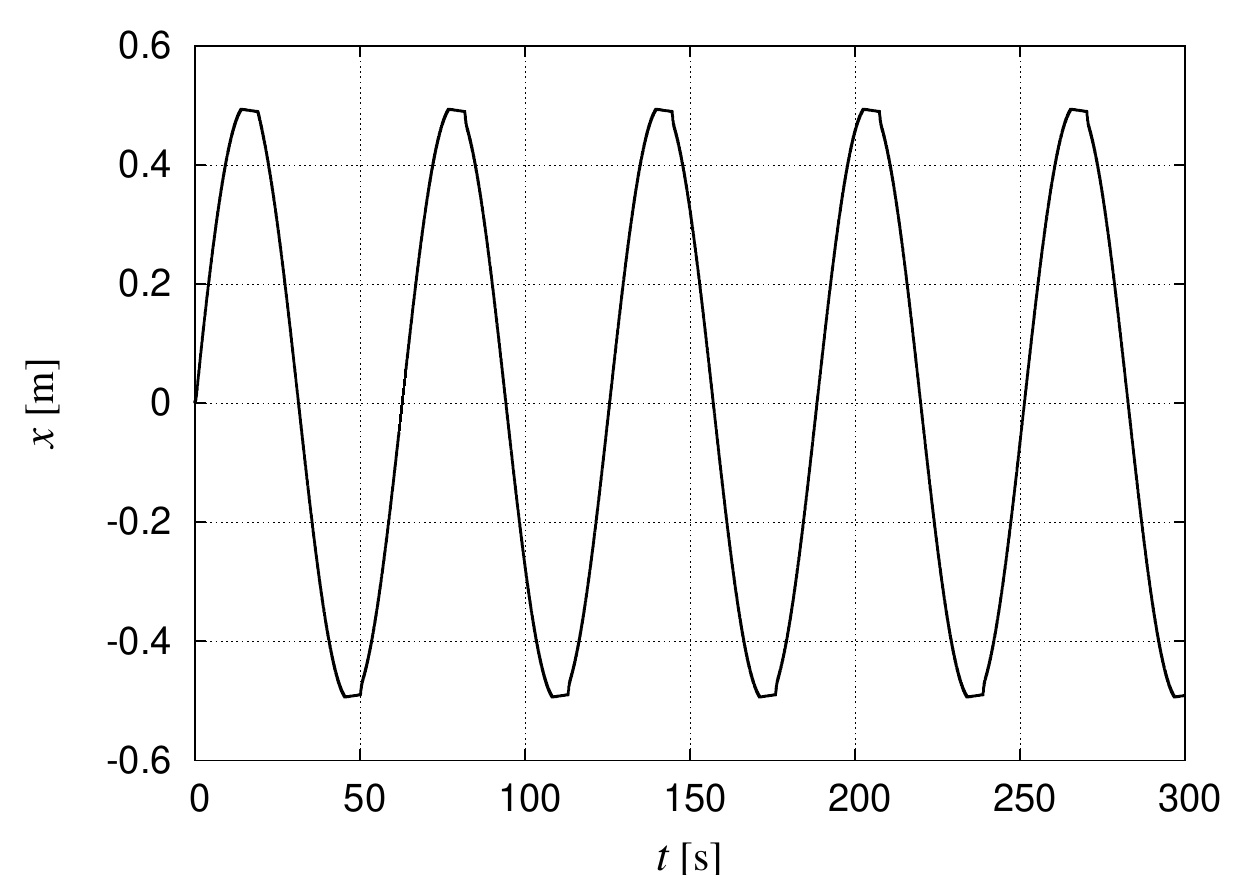}}
\subfigure[Control voltage.]{\label{fig:graf2_1} 
\includegraphics[width=0.35\textwidth]{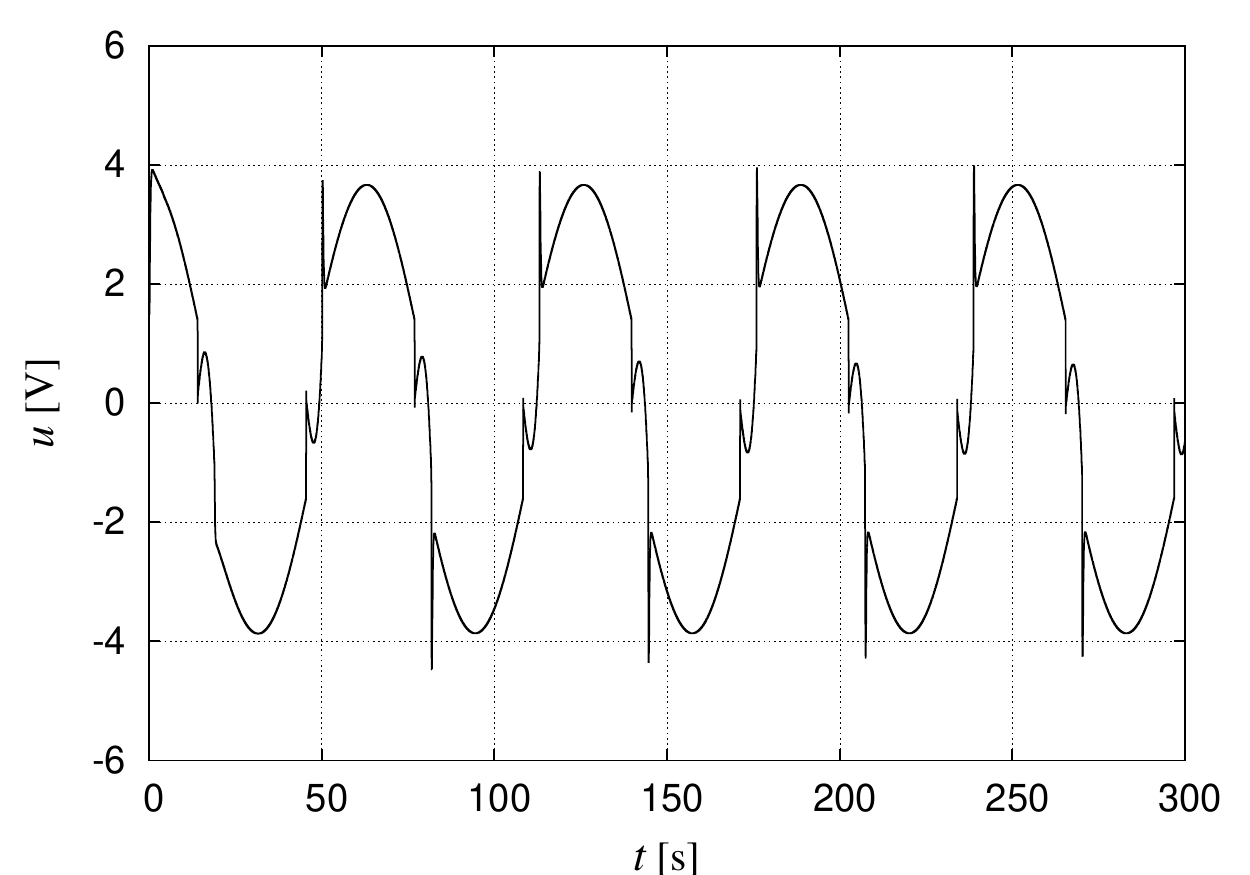}}
}
\mbox{
\subfigure[Tracking error.]{\label{fig:graf3_1} 
\includegraphics[width=0.35\textwidth]{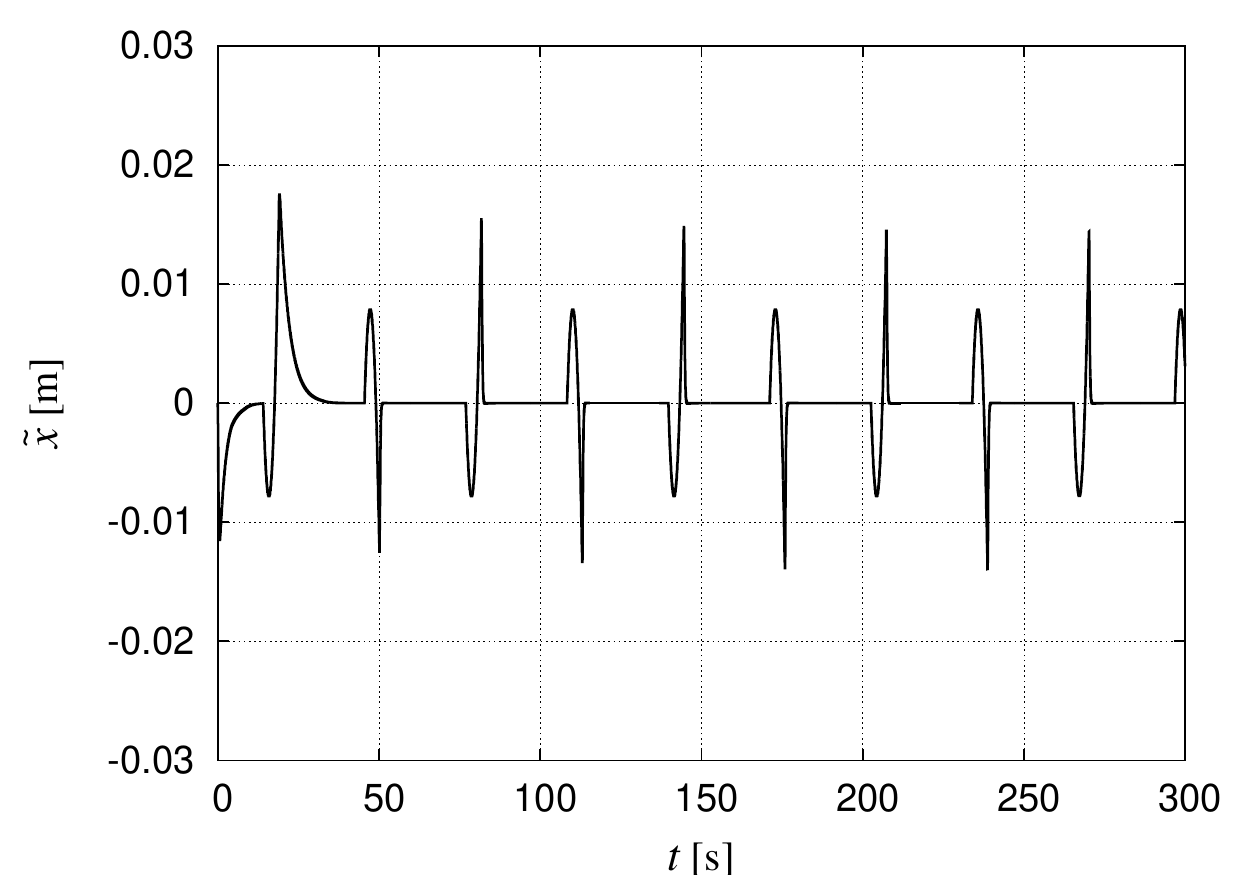}}
\subfigure[Convergence of $\hat{d}(\hat{u})$ to $d(u)$.]{\label{fig:graf4_1} 
\includegraphics[width=0.35\textwidth]{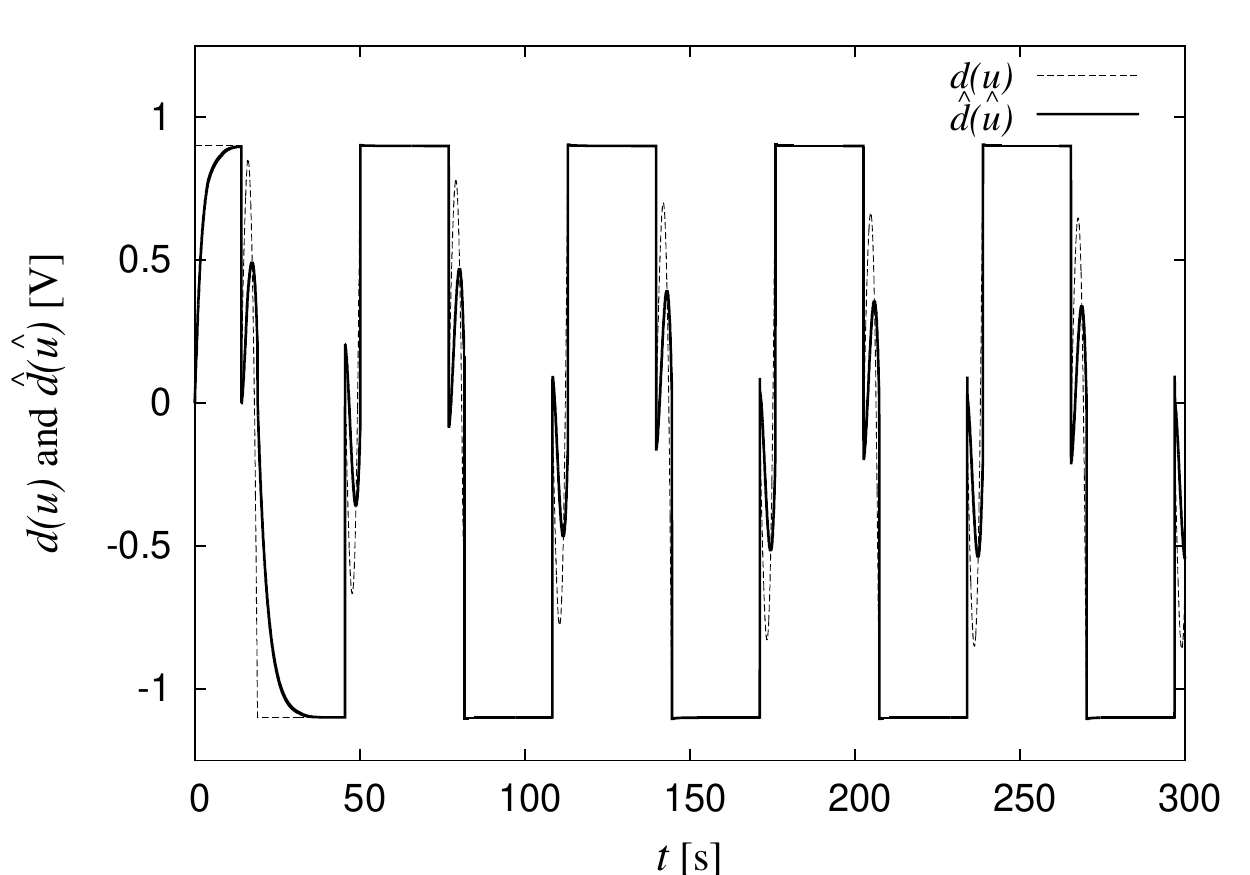}}
}
\caption{Tracking performance with $x_d=0.5\sin(0.1t)$ m and constant supply pressure.}
\label{fig:sim1}
\end{figure}

\begin{figure}[htb]
\centering
\mbox{
\subfigure[Tracking performance.]{\label{fig:graf1_2} 
\includegraphics[width=0.35\textwidth]{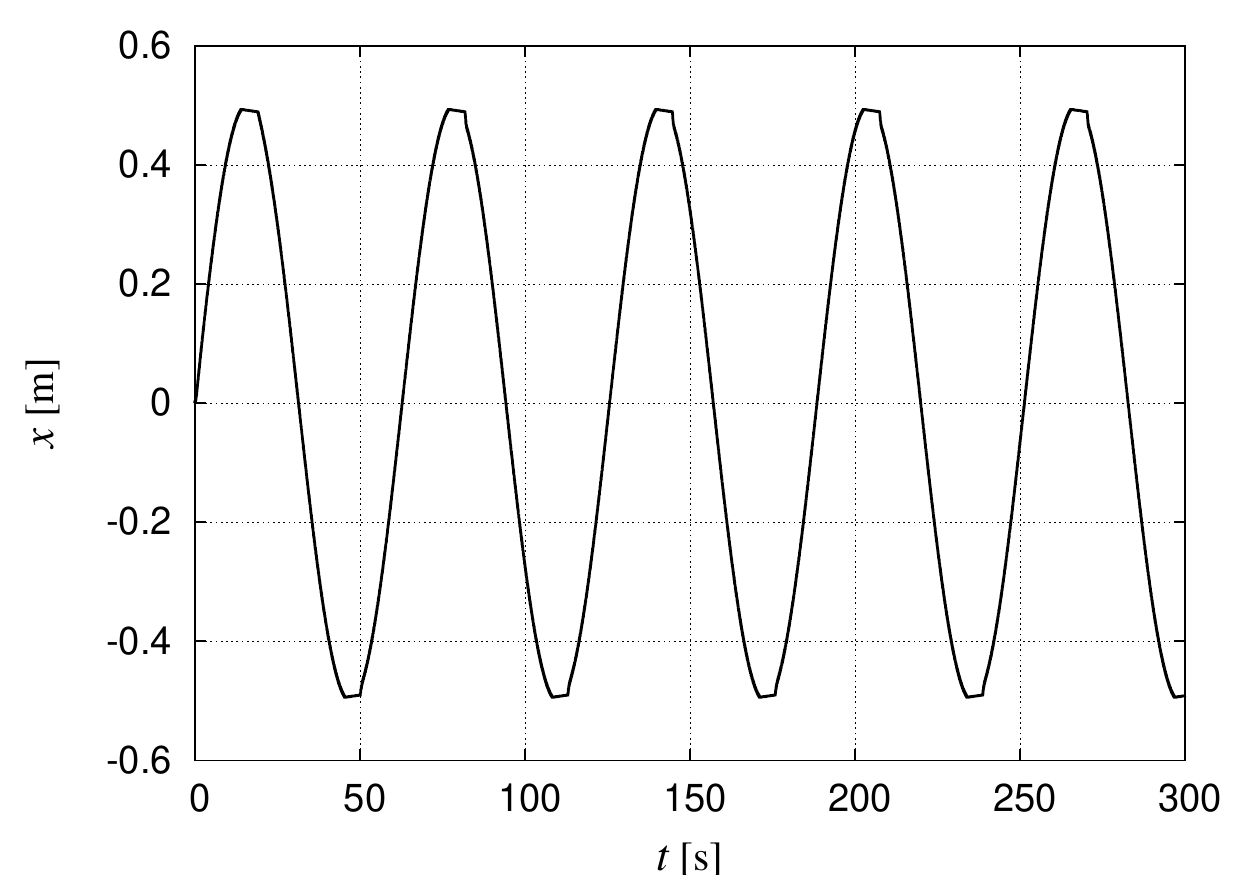}}
\subfigure[Control voltage.]{\label{fig:graf2_2} 
\includegraphics[width=0.35\textwidth]{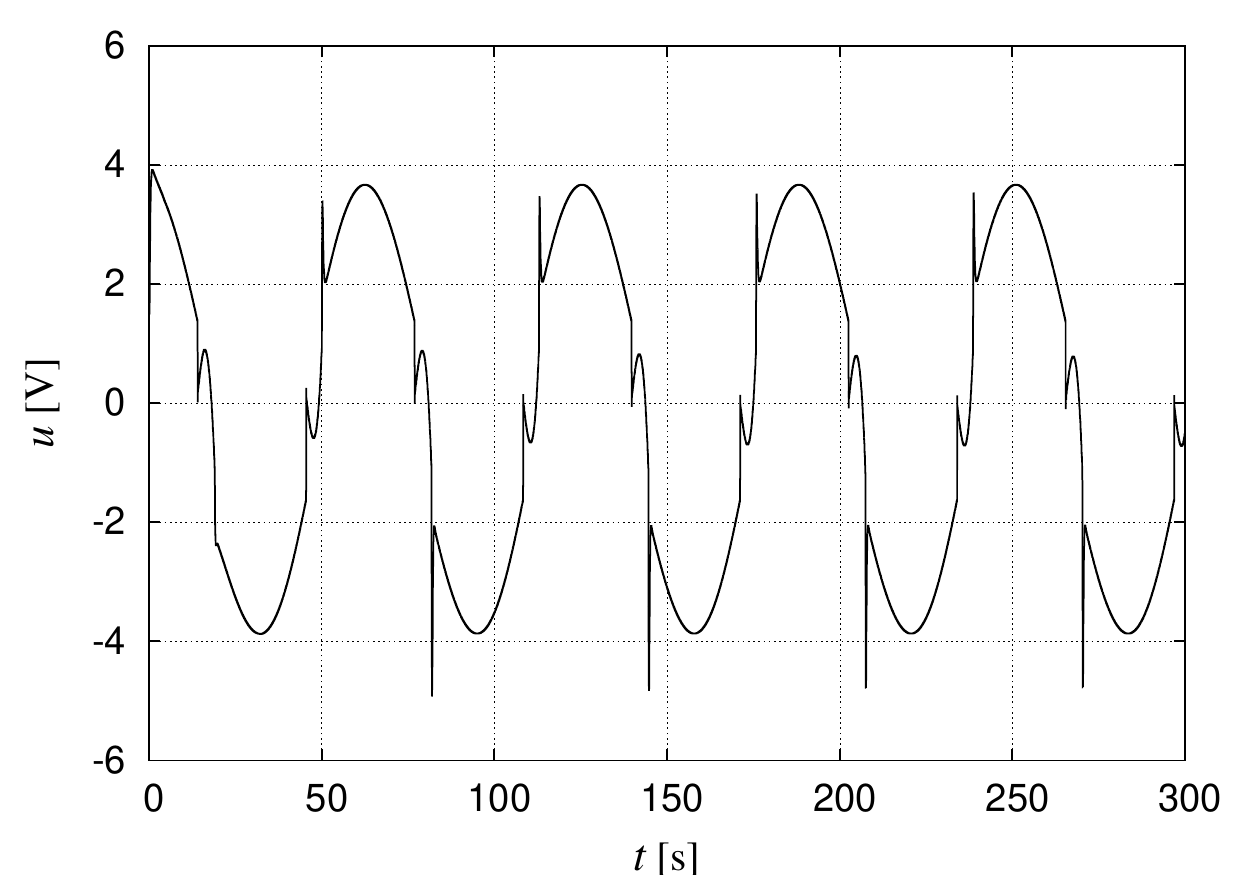}}
}
\mbox{
\subfigure[Tracking error.]{\label{fig:graf3_2} 
\includegraphics[width=0.35\textwidth]{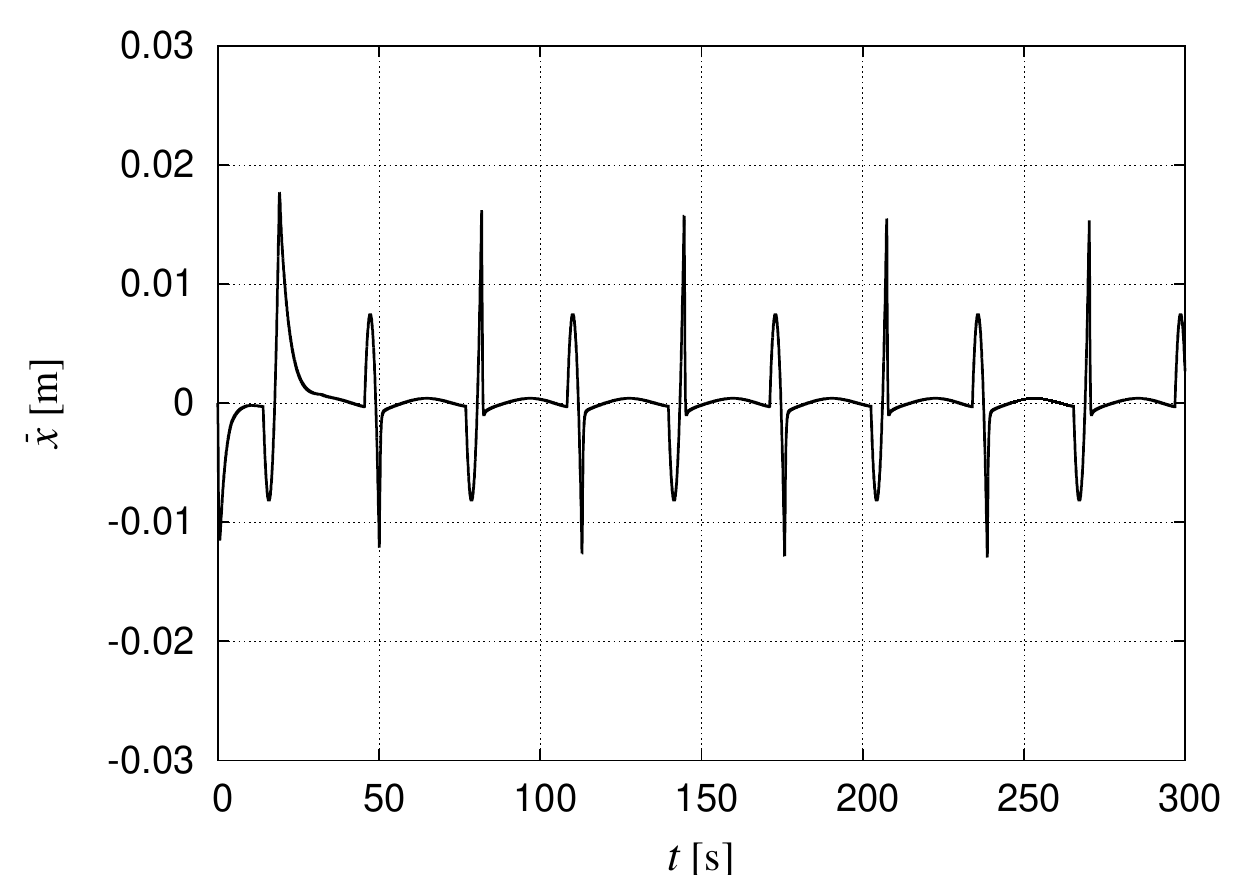}}
\subfigure[Convergence of $\hat{d}(\hat{u})$ to $d(u)$.]{\label{fig:graf4_2} 
\includegraphics[width=0.35\textwidth]{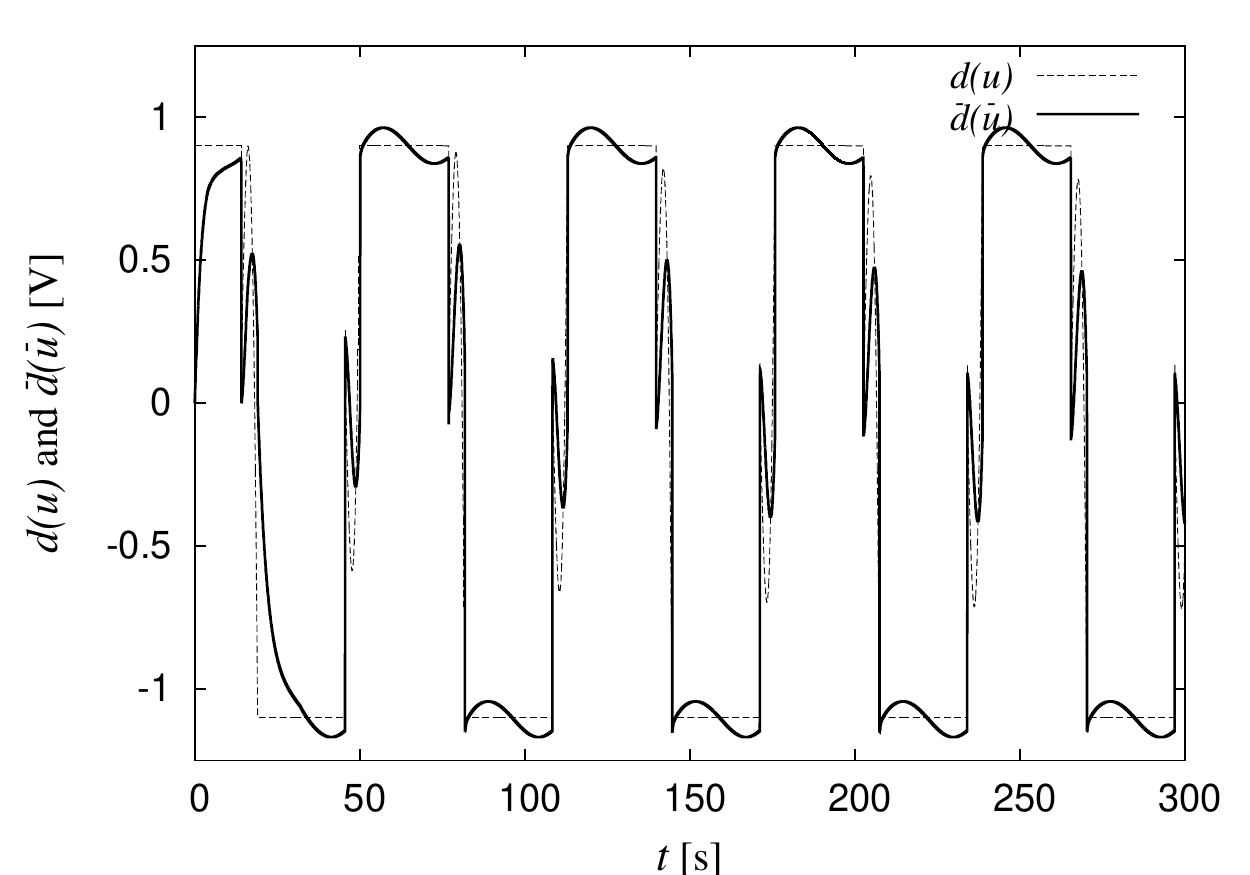}}
}
\caption{Tracking performance with $x_d=0.5\sin(0.1t)$ m and variable supply pressure.}
\label{fig:sim2}
\end{figure}

To evaluate the performance of the proposed control law, Eq.~(\ref{eq:lei}), some numerical simulations 
were carried out. Figure~(\ref{fig:sim1}) shows the results obtained with $x_d=0.5\sin(0.1t)$ m. In 
Fig.~(\ref{fig:sim2}), variations of $\pm20\%$ in the supply pressure, $P_s=7(1+0.2\sin(x))$ MPa, were 
also taken into account. Such variations are very common in real plants.

As observed in Fig.~\ref{fig:graf3_1} and Fig.~\ref{fig:graf3_2}, the adopted controller provides good
tracking performance and is almost indifferent to variations in the supply pressure. It can be easily
verified in Fig.~\ref{fig:graf4_1} and Fig.~\ref{fig:graf4_2}, that, in both cases, the chosen adaptive 
algorithm shows a fast response.

\section{Concluding remarks}

The present work addressed the problem of controlling electrohydraulic servosystems with unknown dead-zone. 
An adaptive fuzzy controller was implemented to deal with the position trajectory tracking problem. The 
stability and convergence properties of the closed-loop systems was proven using Lyapunov stability theory 
and Barbalat's lemma. The control system performance was also confirmed by means of numerical simulations, 
The adaptive algorithm could automatically recognize the dead-zone nonlinearity and previously compensate 
its undesirable effects.

\end{document}